\documentclass[prd,twocolumn,superscriptaddress,amsfonts,amssymb,amsmath,showpacs,showkeys]{revtex4-1}

\usepackage{graphics}
\usepackage{eurosym}
\usepackage{amsfonts}
\usepackage{amssymb}
\usepackage{amsmath}
\usepackage{graphicx}
\usepackage[font={footnotesize,it}]{caption}
\usepackage{hyperref}
\usepackage{float}

\begin{document}

\title{AdS Black Hole with Phantom Scalar Field}
\author{ Limei Zhang}
\affiliation{College of Physics and Space Science, China West Normal University,Nanchong, Sichuan 637002, People's Republic of China}
\author{Xiaoxiong Zeng}
\affiliation{Institute of Theoretical Physics, Chinese Academy of Sciences, Beijing 100190, China
}
\author{ Zhonghua Li}
\email{sclzh888@163.com}
\affiliation{Institute of Theoretical Physics, China West Normal University,Nanchong, Sichuan 637002, People's Republic of China}

\date{\today }

\begin{abstract}
In this paper, we  present an AdS black hole solution with Ricci flat horizon in Einstein-phantom scalar theory.  The phantom scalar fields just depend on the transverse coordinates $x$ and $y$, and  which are parameterized by the parameter $\alpha$. We study the thermodynamics of the AdS phantom black hole. Although its horizon is a Ricci
  flat Euclidean space, we find that the thermodynamical properties of the black hole solution are qualitatively same as  those of  AdS Schwarzschild black hole. Namely there exists
  a minimal temperature, the large  black hole is thermodynamically stable , while the smaller one is  unstable, so there is a so-called Hawking-Page phase transition
  between the large black hole and the thermal gas solution in the AdS spacetime in Poincare coordinates.   We also calculate the entanglement entropy for a strip geometry dual to the
  AdS phantom black holes and find that the behavior of the entanglement entropy is qualitatively the same as that of the black hole thermodynamical entropy.

\end{abstract}

\maketitle

\section{Introduction}

Due to its confined boundary of anti-de Sitter (AdS) spacetime,  thermodynamics of  black hole in AdS space and those in flat or de Sitter spacetimes are  very different.
 For example, for a Schwarzschild  AdS  black hole , it is stable  if its horizon is larger than a certain value, while for a small black hole, the effect of
 cosmological constant can be neglected, the black hole behaves just like a Schwarzschild black hole in flat space ,and it is thermodynamically unstable.  In addition, it is found
 that there is an  AdS black hole only when the temperature of   black hole is larger than a critical value,  while there is only a thermal gas solution when its temperature is less than
 the  value.  The first order phase transition which occurs between the black hole solution and the thermal gas solution in the AdS spacetime is named Hawking-Page phase transition~\cite{HP}.  The phase transition gets its interpretation  following the AdS/CFT correspondence~\cite{Maldacena:1997re,Gubser:1998bc,Witten:1998qj,Witten:1998zw}.

Another big difference between black holes in AdS space and  flat or de Sitter spacetime is that the horizon  of black holes in flat or de Sitter space
must have a spherical structure, while the horizon topology of black holes in AdS spacetime could be a zero, or negative constant curvature surface, except for the case with a positive constant
curvature surface.  These kinds of black holes  with zero or negative constant curvature horizon have been studied in the literature~\cite{Lemos:1994fn}-\cite{Cai:2001dz} ,they are usually called topological AdS black holes . Particularly, we find  these so-called topological AdS  black holes are always thermodynamically stable, and there do not exist Hawking-Page phase transitions related to those topological
black holes~\cite{Birmingham:1998nr}.  Of course, we should mention here that if one of spatial coordinates is compacted for the Ricci flat AdS black holes ,
then the AdS spacetime  in Poincare coordinates is no longer a ground state with respect to the black hole solutions, instead the so-called AdS soliton has a more low energy and then there is a Hawking-Page phase transition between the Ricci flat black hole and AdS soliton~\cite{Surya:2001vj,Cai:2007wz,Aste:2016ac}.The Ricci flat black holes mean   zero curvature horizon black holes.

On the other hand, since people discovered the phenomenon of  accelerated expansion of the universe at the end of last century,  there have been a lot of proposals to explain the cause of this phenomenon.  One  way is  so-called phantom dark energy with state equation, $p=\omega \rho$, where the state parameter satisfies  $\omega <-1$, and
$p$ and $\rho$ represent the pressure and energy density of phantom dark energy, respectively.  To realize  phantom dark energy,  one can alter the mark in front of the kinetic term of a scalar field~\cite{Caldwell:1999ew}.  This is called phantom scalar field.  Except for various studies in cosmology,  the influence of
phantom scalar field in black hole physics has also been extensively investigated.  For instance, the destiny of a black hole owing to accretion of phantom dark energy  has been studied in \cite{Babichev:2004yx}.  Thermodynamics of spherically symmetric black holes  and critical phenomenon with phantom Maxwell field and phantom scalar field  have been discussed in \cite{Azreg-Ainou:2014gja}. Very recently, people have studied thermodynamical geometry of AdS charged black holes with spherical horizon in \cite{Quevedo:2016cge}.

In this study , we consider a Einstein gravity theory  with a negative cosmological constant and massless phantom scalar field.  The black hole will have
a Ricci flat horizon and phantom scalar fields only linearly depend on transverse coordinates.  The thermodynamics of  AdS phantom black hole are studied, and we  find that although
 its horizon is Ricci flat and infinitely extended, the thermodynamical properties of the black hole and AdS Schwarzschild spherically symmetric black hole are qualitatively the same  .
In particular,  Hawking-Page phase transition can emerge between  phantom black hole and the thermal gas solution in AdS spacetime in Poincare coordinates.  Further, we will
calculate holographic entanglement entropy of  a strip geometry which is  dual to the AdS phantom black hole  boundary following the proposal made
by Ryu and Takayanaki in \cite{Ryu:2006bv}.  In recent papers on holographic superconductor models, it has been found that entanglement entropy plays a good probe role to reveal phase
structure and phase transition in those systems.  It can indicate the appearance of a new phase and gives the order of phase transition based on the behavior of
entanglement entropy~\cite{Albash:2012pd}-\cite{Li:2013rhw}. Very recently holographic entanglement entropy dual to a spherically symmetric AdS charged black hole has been
calculated by Johnson \cite{Johnson:2013dka}.  It was found that the entanglement entropy can characterize the related phase transition for a  charged AdS black hole.  Further
studies on  holographic entanglement entropy and its relation to phase transition have been done for various AdS black holes in \cite{Zeng:2015tfj,Zeng:2016sei,Zeng:2015wtt,Mahapatra:2016tae}.
For the AdS phantom black hole presented in this paper, we will further show such a relation which keeps valid even for the case with the Ricci flat horizon black hole.

\section {phantom AdS black hole solution}

We first consider a   Einstein-Hilbert  action with   phantom massless scalar fields and a negative cosmological constant as
 \begin{eqnarray}
  S=\frac{1}{16 \pi G}\int d^{4}x\sqrt{-g } [R-2\Lambda+\frac{1}{2}\Sigma_{i=1}^{2} (\partial\phi_i)^{2}],
\label{eq1.1}
\end{eqnarray}
In the above formula,$G$ is the newtonian gravitational constant , $\phi_i  $ stand for massless phantom scalar fields. In four dimensions, the relationship between the negative cosmological constant $\Lambda  $ and  the AdS curvature radius $ l $  can be as follow
\begin{eqnarray}
\Lambda=-\frac{3}{l^2}.
\label{eq1.2}
\end{eqnarray}
The Einstein's field equations corresponding with action (\ref{eq1.1})  read
\begin{eqnarray}
 R_{\mu \nu}-\frac{1}{2} g^{\mu \nu}R+\Lambda g^{\mu \nu}=T_{\mu\nu},
\label{eq1.3}
\end{eqnarray}
where the  energy  momentum tensor is
  \begin{eqnarray}
   T_{\mu\nu}=\Sigma_{i=1}^{2} [-\nabla_{\mu}\phi_i\nabla_{\nu}\phi_i + \frac{1}{2} g_{\mu\nu}   (\nabla \phi_i)^2].
\label{eq1.4}
\end{eqnarray}
The motion equation of the phantom scalar field is
\begin{equation}
\nabla^2 \phi_i=0.
\label{eq1.5}
\end{equation}

Now we consider a black hole solution which has  a Ricci flat horizon in this system  with the metric ansatz as
\begin{eqnarray}
   ds^2=-f(r)dt^2+\frac{1}{f(r)}dr^2+r^2 (dx^2 +dy^2).
\label{eq1.6}
\end{eqnarray}
With this metric, one has the non-zero  part of Ricci tensor as~\cite{Birmingham:1998nr}
\begin{eqnarray}
&&  R_{tt} =\frac{1}{2}ff^{\prime\prime}+\frac{1}{r}ff^{\prime},\nonumber\\
 &&R_{rr } =-\frac{1}{2}\frac{f^{\prime\prime}}{f}-\frac{1}{r}\frac{f^{\prime}}{f},\nonumber\\
 && R_{xx} =R_{yy}=-  (f+r f^{\prime}),
\label{eq1.7}
\end{eqnarray}
where  $ f^{\prime}={df}/{dr}$, and $f^{\prime\prime}={d^{2}f}/{d^{2}r}  $. The  Ricci scalar can be written
\begin{eqnarray}
  R=g^{tt}R_{tt}+g^{rr}R_{rr}+g^{i j}R_{i j}=-f^{\prime\prime}-\frac{4}{r} f^{\prime}-\frac{2 f}{r^{2}},
\label{eq1.8}
\end{eqnarray}
We  take the scalar fields  to be functions of   transverse coordinates as
 \begin{eqnarray}
   \phi_{1}=\alpha x , \ \ \  \phi_{2}=\alpha y
\label{eq1.9}
\end{eqnarray}
 where $\alpha$ is a constant, then it is found that the equations of motions for both gravitational field and matter field are satisfied if   $f(r)$ takes the form
\begin{eqnarray}
  f(r)=\alpha^{2}+\frac{r^{2}}{l^{2}}-\frac{2M}{r},
\label{eq1.10}
\end{eqnarray}
in which $M$ is a parameter relating to the mass of the solution.  In fact,  this solution describes an AdS black hole with a Ricci flat horizon . As we can see that
the influence of the phantom scalar in the solution is manifested by the parameter $\alpha$.  In addition,  note that the scalar field enters the action only through its derivative and therefore enjoys a shift symmetry. These scalar fields are only determined up to a constant.   When the phantom scalar fields are absent, the solution reduces to the AdS
black hole solution with Ricci flat horizon in Einstein gravity. In that case, one has $\alpha=0$.  A natural generalization of the solution (\ref{eq1.10}) is to include electric and magnetic
charged from Maxwell field.  Namely if a Maxwell field appears in  action (\ref{eq1.1}),  then one has a dynamic phantom AdS black hole solution  as
\begin{equation}
f(r) = \alpha^2 +\frac{r^2}{l^2} -\frac{2M}{r} + \frac{e^2+g^2}{r^2},
\label{eq1.11}
\end{equation}
where $e$ and $g$ are electric charge and magnetic charge of the solution, respectively.  Here  we mention that one can easily generalize these black hole solutions
(\ref{eq1.10}) and (\ref{eq1.11}) to higher dimensional ($d>4$) cases.  Finally we should stress that if the scalar fields are the usual canonical ones in (\ref{eq1.1}), namely the
sign of the kinetic term in  the scalar fields formula  is negative, this AdS black hole  solution was first found in \cite{Bardoux:2012aw} with the metric function
\begin{equation}
f(r) = -\alpha^2 +\frac{r^2}{l^2}-\frac{2M}{r}.
\label{eq1.12}
\end{equation}
In that case, the thermodynamical properties of the solution (\ref{eq1.12}) and the topological black hole  with a horizon of negative constant
curvature space are qualitatively same in AdS spacetime~\cite{Birmingham:1998nr}.

\section{Thermodynamics of  the phantom AdS black hole}

In this section,  the thermodynamics  of  this  AdS black hole solution are discussed.  Here we are discussing the case  where
 the transverse coordinates $x$ and $y$ are infinitely extended. In other words, the directions spanned by $x$ and $y$ are not compacted. In that case, precisely speaking,
 what we are discussing is not a black hole, but a black brane.  And the vacuum of the system is no longer the AdS solution, but an AdS space in Poincare coordinates with
 metric function
 \begin{equation}
 \label{vacuum}
 f(r)= \frac{r^2}{l^2}.
 \end{equation}
On the other hand, let us note that when $\alpha^2=1$,   the metric function in (\ref{eq1.10}) is nothing, but the one for a Schwarzschild AdS black hole with a spherical horizon.  Thus one may expect that the thermodynamic properties of the black hole solution (\ref{eq1.10}) are essentially the same  as those of AdS Schwarzschild
black holes.  Here we will show this indeed holds.  To make qualitative comparison, we take the area spanned by $x$ and $y$ is $4\pi$. So  the parameter $M$ in (\ref{eq1.10})
is just the mass of the solution with respect to the AdS vacuum in Poincare coordinates.  In this paper we take the units with $G=c=h=k_b=1$.

 The horizon of the black hole is given by $ f(r)\mid_{r=r_{+}}=0 $  . Thus the  black hole  mass can be expressed in terms of the horizon radius $ r_{+} $ as

 \begin{eqnarray}
   M=\frac{\alpha^2}{2}r_{+}+\frac{r_{+}^3}{2l^2}.
\label{eq2.1}
\end{eqnarray}
As a function of the horizon radius $ r_{+} $, $ M  $ is a monotonically increasing function. In other words, this equation shows that for any positive value of $ M  $ there is a corresponding horizon. And only when the case  $ r_{+} \rightarrow 0$, $ M\rightarrow 0 $.
The Hawking temperature $ T_{H} $ of the black hole can be obtained in several ways.  One way is that  the Euclidean time of the black hole solution should take a period so that
the potential conical singularity can be avoided at the horizon of the black hole and the inverse of the period just gives the Hawking temperature. This way, one has the Hawking
temperature of the black hole as
  \begin{eqnarray}
   T=\frac{1}{4\pi r_{+}}\left (\alpha^{2}+\frac{3 r_{+}^2}{l^2}\right ).
\label{eq2.2}
\end{eqnarray}
In Einstein's gravity theory,  the entropy of black hole obeys the well-known Bekenstein-Hawking area formula.  In our case, the entropy of  black hole can be expressed
\begin{eqnarray}
   S=\pi r_{+}^2.
\label{eq2.3}
\end{eqnarray}
Substituting (16) into (15) and eliminating  $ r_{+} $,  $ T  $ is given by
\begin{eqnarray}
   T=\frac{\alpha^{2}+\frac{3 S}{\pi l^{2}}}{4\sqrt{\pi} \sqrt{S}}.
\label{eq2.4}
\end{eqnarray}
It is easy to show that the mass, entropy and the temperature of the black hole obey the first law of black hole thermodynamics
\begin{eqnarray}
   d M=T d S.
\label{eq2.5}
\end{eqnarray}
To study the thermodynamical stability of the black hole, let us calculate  the heat capacity,
\begin{eqnarray}
   C=T\frac{\partial S}{\partial T}=\frac{\partial M} {\partial T}=\frac{2 l^2 \pi r_{+}^2 (\frac{3 r_{+}^2}{l^2}+\alpha^2)}{3 r_{+}^2-l^2 \alpha^2}.
\label{eq2.6}
\end{eqnarray}
We can see that the heat capacity is positive when  $r_+ > l\alpha/\sqrt{3}$,   negative  when $r_+ <l\alpha/\sqrt{3}$,  and divergent when $r_+=l\alpha/\sqrt{3}$.
This means that the black hole is locally thermodynamic unstable for small horizons with  $r_+<l\alpha/\sqrt{3}$, while it is locally thermodynamic stable for large
horizons with  $r_+ > l\alpha/\sqrt{3}$.     Figure 1 shows the temperature-entropy relation .  And it is easy to see that  the small entropy  is thermodynamically unstable, while the large entropy is thermodynamically stable.
\begin{figure}[H]
\centering
\includegraphics[width=0.45\columnwidth]{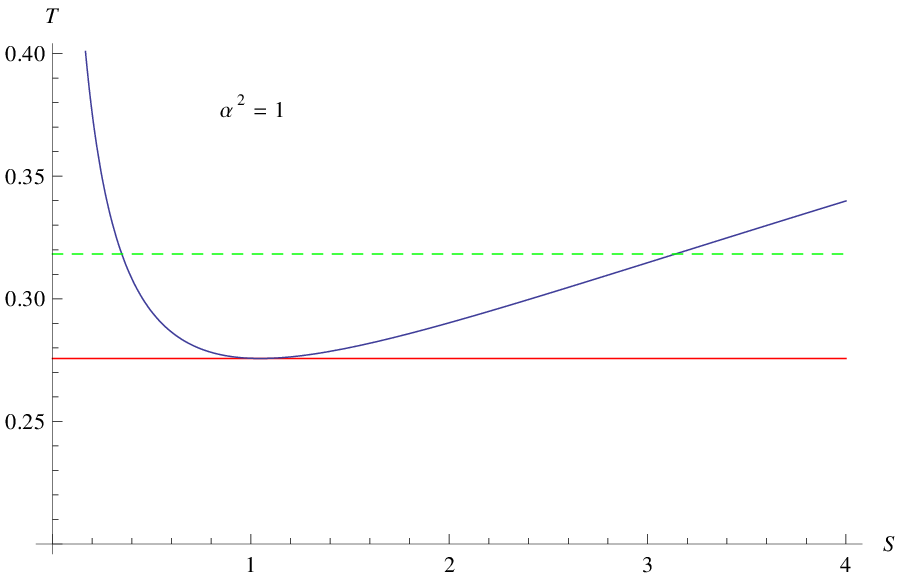}
\includegraphics[width=0.45\columnwidth]{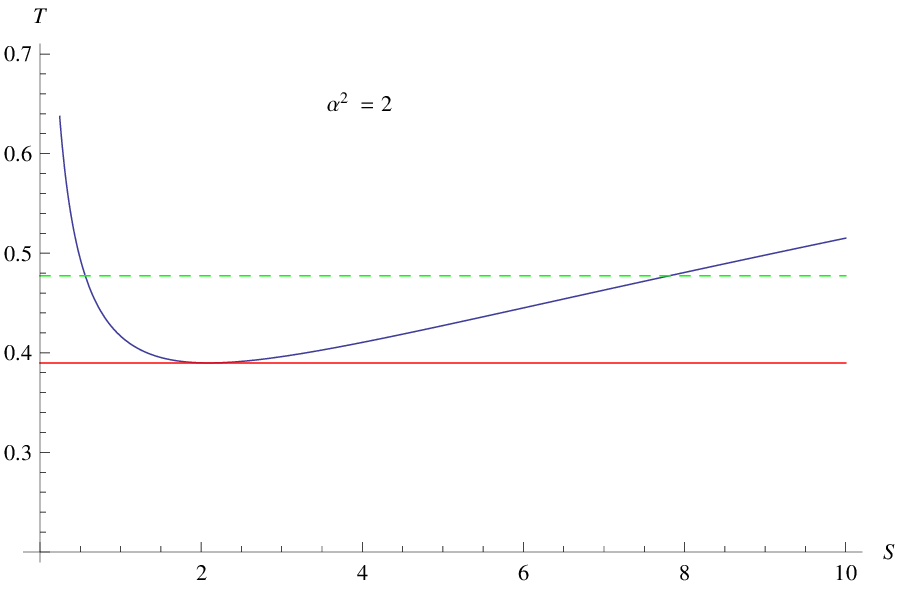}
\caption{The black hole temperature in term of entropy.  The dashed green  line represents the  Hawking-Page  transition temperature, while the solid  red line stands for   $T_{\rm min}$  . When $ T<T_{\rm min}  $, there is only  a thermal gas  solution  in the AdS space.  When  $ T>T_{\rm min}  $, for a given temperature  $ T $,  there exist two black hole solutions given by  (\ref{eq3.4}).  The one with large horizon is  stable, while it is thermodynamically
unstable for the smaller one .}
\end{figure}

The Hawking temperature (\ref{eq2.2}) can be re-expressed as
 \begin{eqnarray}
 \frac{3}{l^2} r_{+}^{2}-4\pi k _{B}T r_{+}+\alpha^{2}=0.
\label{eq3.1}
\end{eqnarray}
Through this relation, we can see that there exists a minimal temperature as
\begin{eqnarray}
  T _{\rm min} =\frac{\alpha \sqrt{3}}{2 \pi l}.
\label{eq3.2}
\end{eqnarray}
 In that case, corresponding horizon radius is given by
 \begin{eqnarray}
 r _{0}=\frac{\alpha l }{\sqrt{3}}.
\label{eq3.3}
\end{eqnarray}
When $ T>T_{\rm min} $, for a given temperature there are two black hole horizons
\begin{eqnarray}
r_{l,s}=\frac{\alpha^2 T }{2\pi T_{min}^{2}}(1+\sqrt{1-\frac{ T_{min}^{2}}{T^2}}).
\label{eq3.4}
\end{eqnarray}
The  small black hole with negative heat capacity  is thermodynamical unstable , while the larger one  with positive heat capacity is thermodynamical stable.

With the mass, entropy and temperature, the Helmholtz free energy  is easily calculated as
\begin{eqnarray}
F=M-TS=\frac{1}{4} r_+ (-\frac{r^2_+}{l^2} + \alpha ^2).
\label{eq3.5}
\end{eqnarray}
 According to the above ,we can easily see  that the free energy is equivalent to the Euclidean action  by setting the pure AdS vacuum in Poincare coordinates as a ground state.  We can see
from the free energy that the free energy is positive when  $r_+ <l \alpha$,  while it is negative when $r_+ >l\alpha$.  It implies that the black hole is globally thermodynamic unstable
when $r_+ <l\alpha$, while it is globally thermodynamic stable when $r_+> l\alpha$.  Namely there is a Hawking-Page phase transition   when the free energy vanishes and $r_+=l\alpha$. The  Hawking-Page  transition temperature is given by
\begin{eqnarray}
T _{\rm HP}=\frac{\alpha}{\pi l}.
\label{eq3.6}
\end{eqnarray}
When $T>T_{\rm HP}$,  the system is dominated by the black hole solution, while it is dominated by a thermal gas solution as $T<T_{\rm HP}$.  When $T=T_{\rm HP}$,  the
Hawking-Page transition happens between the black hole phase and thermal gas phase.

From (\ref{eq3.2}) and (\ref{eq3.6}), we can see that  the phantom scalar field  has some effect on the  thermodynamic stability and the Hawking-Page transition temperature through
the parameter $\alpha$. But  the ratio between these two temperatures
  \begin{eqnarray}
D=\frac{T_{\rm HP}}{T_{\rm min}}=\frac{2\sqrt{3}}{3},
\label{eq3.7}
\end{eqnarray}
is independent of the parameter $\alpha$.  The ratio is the same as the one for the case of  the AdS Schwarzschild black hole.  We show the free energy of the
black hole as a function of temperature in Figure 2 .

\begin{figure}[H]
\centering
\includegraphics[width=0.45\columnwidth]{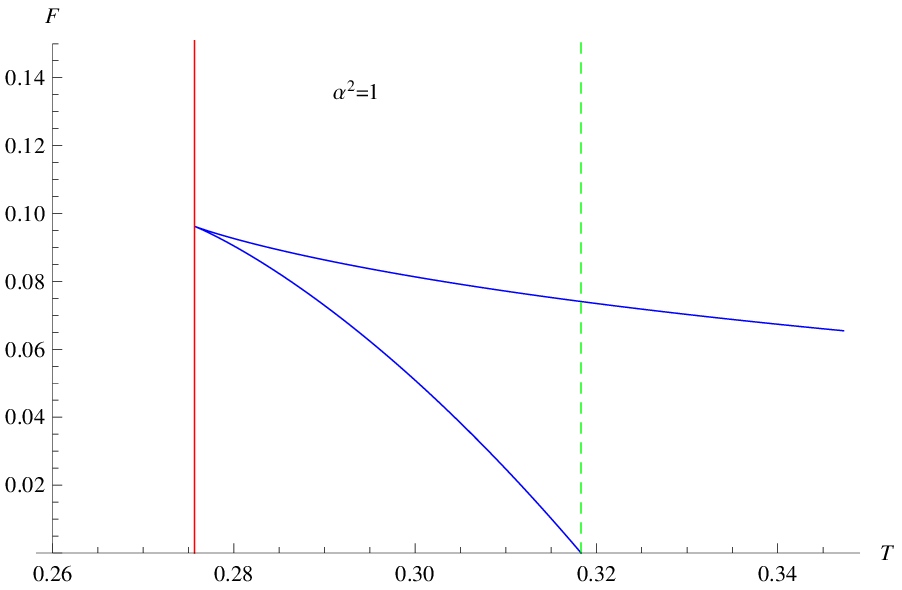}
\includegraphics[width=0.45\columnwidth]{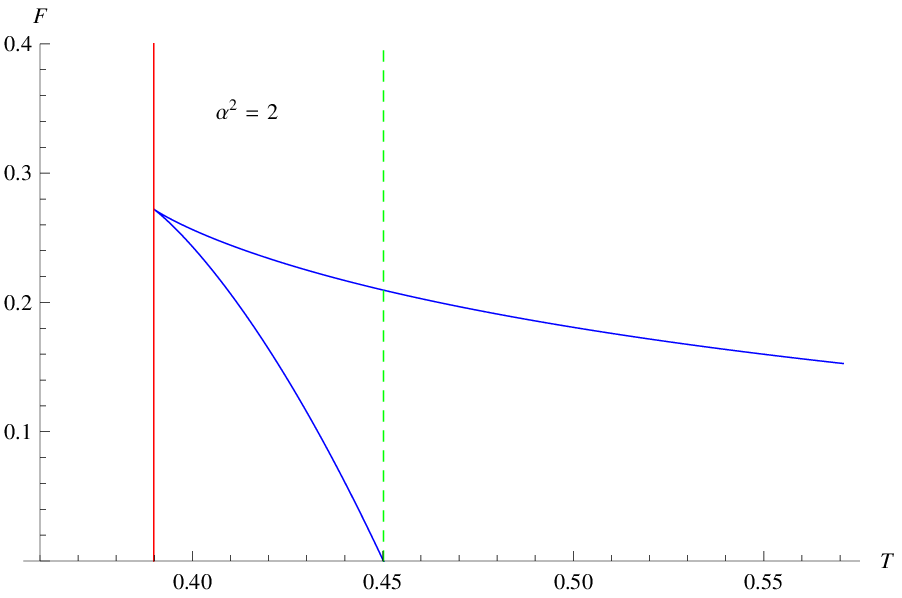}
\caption{The Helmholtz free energy  of the black hole as a function of  temperature $T  $.  The dashed green line stands for the Hawking-Page  transition temperature, while
the solid red  line represents  the minimal  temperature.}
\end{figure}
\section{Entanglement entropy and Hawking-Page transition}
\label{sec:3}
For a quantum mechanical system with many degrees of freedom, we separate  the total system into  two subsystems $A$ and $B$. The Hilbert space of the  total system can be expressed by the direct product of the two subsystems  $A$ and $B$ as  $H_{\rm total}=H_{A}\bigotimes H_{B}  $. For the observer who is only accessible to   the  subsystem $A$, because the information of $B$ can not be directly observed, the total system is described by reduced density matrix $ \rho_{A} $,
\begin{eqnarray}
\rho_{A}=\rm {tr}_{B}  \rho_{\rm total},
\label{eq4.1}
\end{eqnarray}
where  the trace is taken only over the Hilbert space of subsystem $B$. The entanglement entropy of the subsystem $A$ is defined   as
\begin{eqnarray}
S_{A}=-{\rm tr}_{A}\rho_{A} {\rm  log} {\rho_{A}}.
\label{eq4.2}
\end{eqnarray}
Usually it is very difficult to calculate entanglement entropy  in  quantum field theory.   However, the AdS/CFT correspondence provides  a powerful tool to calculate the entanglement entropy  for a strong coupling conformal field theory in a simple way.   Suppose the boundary of the subsystem $A$ is $ \partial A $ in the boundary of the AdS spacetime. The entanglement  entropy of the subsystem $A$ is given by the following formula~\cite{Ryu:2006bv}
\begin{eqnarray}
S_{A}=\frac{{\rm Area} (\gamma_{A})}{4 G},
\label{eq4.3}
\end{eqnarray}
 where $ G $ is the Newton gravitational  constant in the AdS bulk, $ \gamma $ is the minimal surface extended into the bulk with boundary  $ \partial\gamma=\partial A $.

 Now we will take the formula (\ref{eq4.3})  to calculate the entanglement entropy of the conformal field theory dual to the AdS black hole solution given by (\ref{eq1.10}). To study the entanglement entropy, we should choose a proper region for A. Here we consider a rectangular  strip parameterized  by the boundary coordinates $ x  $ and $y  $, and we assume  $ y $-direction is infinitely extended and the direction $x$ has
 a width $\ell$. Thus  we can  use   $  x $ to parameterize the minimal surface.

With the help of  (\ref{eq1.6}) and (\ref{eq4.3}), we can obtain the area of the minimal surface for the strip geometry as
  \begin{eqnarray}
  {\rm Area} (\gamma_A) =L\int_{-{\ell}/{2}} ^{{\ell}/{2}}dx \sqrt{\frac{(r^{\prime})^2}{f(r)} +r} ,
\label{eq4.4}
\end{eqnarray}
where $r^{\prime}= {dr}/{dx}  $ and $  L  $ is the length  along  $ y $-direction, which will be set to be unity in what follows. Note  that the integrand in  Eq.~(\ref{eq4.4}) does not depend explicitly
 on $ x $.  We can derive  the equation of motion of $ r(x) $ as,
  \begin{eqnarray}
 4 r(x)^{2} f(r)^{2}-2 f(r) r^{\prime}(x)^{2} \nonumber\\
 +r(x)r^{\prime}(x)^{2}f^{\prime}(r)-2 f(r) r r^{\prime\prime} (x)=0.
\label{eq4.5}
\end{eqnarray}
Due to the symmetry of the minimal surface,  obviously we have
 \begin{eqnarray}
  r(0)=r_{0},\ \ \ \ r^{\prime}(0)=0,
\label{eq4.6}
\end{eqnarray}
at the returning point $r=r_0$ where $x=0$.  With the condition (\ref{eq4.6}), we can solve Eq.~(\ref{eq4.5})  numerically and obtain the function $r(x)$.   Then substituting $r(x)  $ into (\ref{eq4.4}), we can  get the entanglement entropy.  It is found that  the entanglement entropy is always divergent.  In order to get more  meaningful physically results, we introduce an ultraviolet cut-off.   Here we are interested in the regularized  entropy $ \delta S=S-S_{0} $, where $ S_{0} $ is the entanglement entropy  for the same geometry, but the bulk is the
pure AdS space in Poincare coordinates, which is also calculated with numerical method.  In Figure 3, as typical examples, we plot the regularized entanglement entropy versus the temperature of
the black hole in the cases $ \ell=0.6, \ell=1.2 $ and $ \alpha^{2}=1, \alpha^{2} =2$, the corresponding ultraviolet cut-off is set to be $ r(0.59), r(1.19)$.
Comparing  Figure 3 with Figure 1, we find  that the behavior of the entanglement entropy is quite similar to that  of the black hole entropy,  and the corresponding minimal temperature  $  T_{\rm min} $,  which is showed by solid red line in  Figure 3,   are the same as  those in  Figure 1.

In Figure 3 we also show the Hawking-Page transition temperature  with dashed green line. With  (\ref{eq3.7}),  it shows that  the ratio between the minimal temperature and Hawking-Page transition temperature  also keeps valid in the behavior of the entanglement entropy.    In addition, we notice that with the change of the width $\ell$, the entanglement entropy
changes consequently, but the phase structure of the entanglement entropy does not changes and the minimal temperature does not change.  We can therefore conclude that the entanglement entropy can indeed reveal the phase structure of the AdS phantom black hole.

\begin{figure}[H]
\centering
\includegraphics[width=0.45\columnwidth]{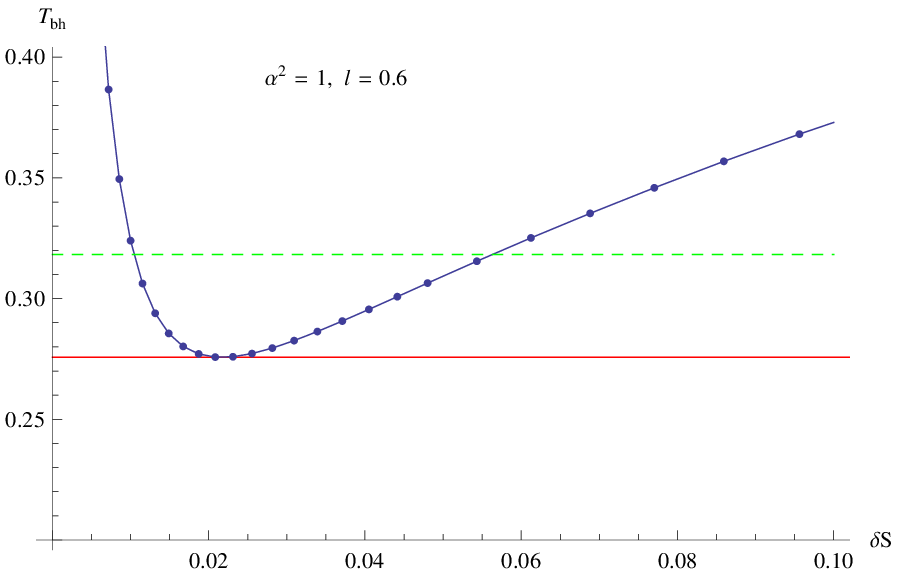}
\includegraphics[width=0.45\columnwidth]{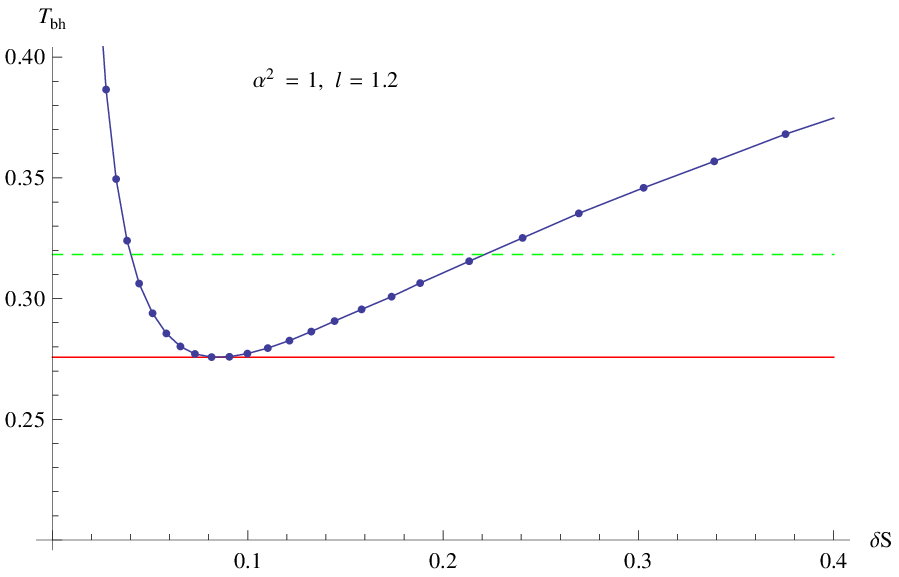}
\includegraphics[width=0.45\columnwidth]{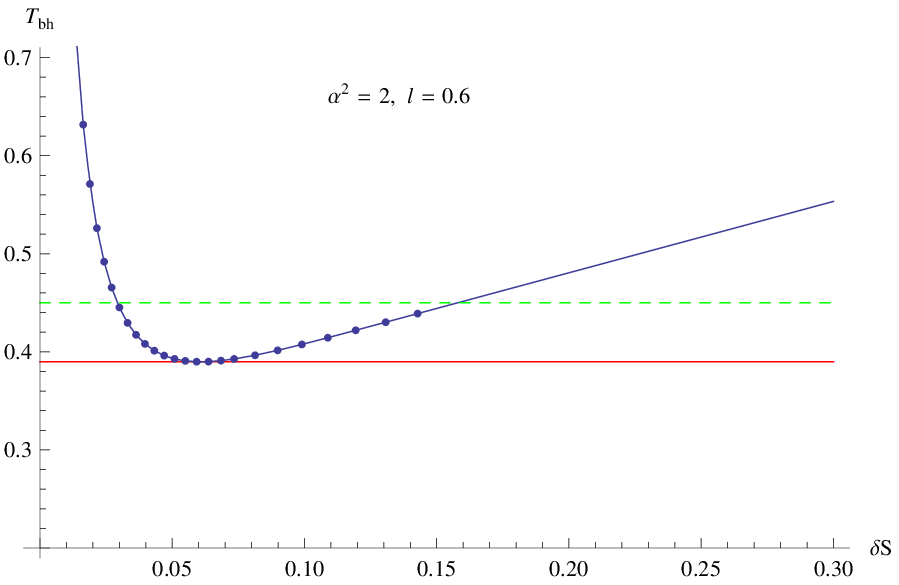}
\includegraphics[width=0.45\columnwidth]{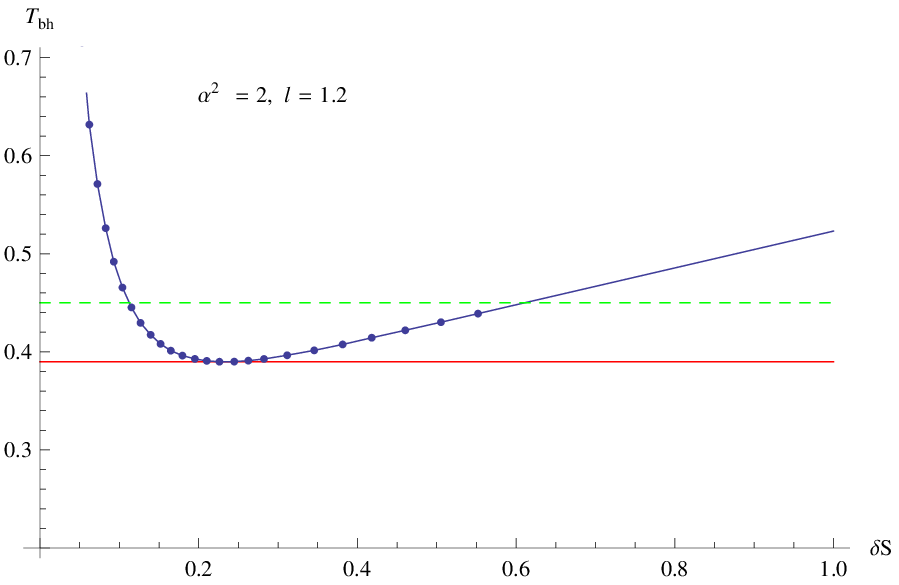}

\caption{The regularized entanglement entropy versus the black hole temperature.  The dashed green line represents the Hawking-Page transition temperature, while the solid red  line stands for the minimal  temperature. }
\end{figure}

\section{Discussion and conclusions}
Black hole physics is one of important topics in general relativity and quantum gravity.  Thermodynamics, quantum mechanics, statistical physics and information theory are
entangled with each other in black hole thermodynamics.  It is generally believed that to completely understand black hole thermodynamics will be greatly helpful to establish
a self-consistent quantum gravity theory.  In this respect, the AdS black hole might play a central role since the AdS/CFT correspondence relates a quantum gravity in AdS spacetime
to a conformal field theory in the AdS boundary.  In this paper we have presented an AdS black hole solution with Ricci flat horizon in Einstein-phantom scalar theory.  The phantom
scalar fields only depend on the transverse coordinates $x$ and $y$, and are parameterized by the parameter $\alpha$ (see Eq.~(\ref{eq1.9})).   However, we noticed that the parameter
$\alpha$ can be renormalized to be one by the following rescaling,
\begin{equation}
\label{eq5.1}
t \to t/\alpha, \ \ \  r \to \alpha r,  \ \ \, (x, y) \to (x,y)/\alpha, \  \ \  M \to \alpha^3 M.
\end{equation}
With the rescaling (\ref{eq5.1}), the black hole solution can be rewritten as
\begin{equation}
\label{eq5.2}
ds^2=-g(r) dt^2 +g^{-1}(r) dr^2 +r^2 (dx^2+dy^2),
\end{equation}
with
$$ g(r)= 1 +\frac{r^2}{l^2} -\frac{2M}{r}.$$
Comparing this black hole solution with the AdS Schwarzschild black hole
\begin{equation}
ds^2=-g(r) dt^2 +g^{-1}(r) dr^2 + r^2 (d\theta^2 +\sin^2\theta d\varphi^2),
\label{eq5.3}
\end{equation}
we note that the only difference is the replacement of  the 2-dimensional Euclidean space $R^2$ by the 2-dimensional sphere $S^2$. 

Indeed we have found that although its horizon structure is Ricci flat, the AdS black hole with phantom scalar has qualitatively same thermodynamic properties as those
of AdS Schwarzschild black hole: there exists a minimal temperature, small black hole is thermodynamically unstable, while large black hole is thermodynamically stable.
And in particular, the Hawking-Page phase transition can happen between the AdS phantom black hole and the thermal gas solution in AdS spacetime in Poincare coordinate.

Further we have calculated the entanglement entropy of dual conformal field theory for a strip geometry in the AdS boundary by the Ryu-Takayanaki proposal.  It was found that
the behavior of the entanglement entropy is also qualitatively same as that of the thermodynamical entropy of the black hole.  This example further supports that the black hole entropy
might indeed be interpreted as entanglement entropy.  On the other hand, it also indicates  entanglement entropy is indeed a good probe to phase structure and phase transition
in black hole thermodynamics.

\begin{acknowledgments}

This work is supported by the Meritocracy Research Funds of China West Normal University and the
authors thank R.G. Cai for helpful suggestions and discussions.

\end{acknowledgments}

\end{document}